\PassOptionsToPackage{unicode}{hyperref}
\PassOptionsToPackage{hyphens}{url}
\documentclass[
]{article}
\usepackage{amsmath,amssymb}
\usepackage{lmodern}
\usepackage{iftex}
\ifPDFTeX
  \usepackage[T1]{fontenc}
  \usepackage[utf8]{inputenc}
  \usepackage{textcomp} % provide euro and other symbols
\else % if luatex or xetex
  \usepackage{unicode-math}
  \defaultfontfeatures{Scale=MatchLowercase}
  \defaultfontfeatures[\rmfamily]{Ligatures=TeX,Scale=1}
\fi
\IfFileExists{upquote.sty}{\usepackage{upquote}}{}
\IfFileExists{microtype.sty}{% use microtype if available
  \usepackage[]{microtype}
  \UseMicrotypeSet[protrusion]{basicmath} % disable protrusion for tt fonts
}{}
\makeatletter
\@ifundefined{KOMAClassName}{% if non-KOMA class
  \IfFileExists{parskip.sty}{%
    \usepackage{parskip}
  }{% else
    \setlength{\parindent}{0pt}
    \setlength{\parskip}{6pt plus 2pt minus 1pt}}
}{% if KOMA class
  \KOMAoptions{parskip=half}}
\makeatother
\usepackage{xcolor}
\IfFileExists{xurl.sty}{\usepackage{xurl}}{} % add URL line breaks if available
\IfFileExists{bookmark.sty}{\usepackage{bookmark}}{\usepackage{hyperref}}
\hypersetup{
  hidelinks,
  pdfcreator={LaTeX via pandoc}}
\usepackage{graphicx}
\makeatletter
\def\maxwidth{\ifdim\Gin@nat@width>\linewidth\linewidth\else\Gin@nat@width\fi}
\def\maxheight{\ifdim\Gin@nat@height>\textheight\textheight\else\Gin@nat@height\fi}
\makeatother
\setkeys{Gin}{width=\maxwidth,height=\maxheight,keepaspectratio}
\makeatletter
\def\fps@figure{htbp}
\makeatother
\ifLuaTeX
  \usepackage{selnolig}  % disable illegal ligatures
\fi

\author{}
\date{}

\begin{document}

\textbf{A Large-Scale Nanocrystal Database with Aligned Synthesis and
Properties Enabling Generative Inverse Design}

Kai Gu\textsuperscript{1}, Yingping Liang\textsuperscript{2}, Senliang
Peng\textsuperscript{1}, Aotian Guo\textsuperscript{1}, Ying
Fu\textsuperscript{2,*}, Haizheng Zhong\textsuperscript{1,*}

\textsuperscript{1}MIIT Key Laboratory for Low-Dimensional Quantum
Structure and Devices, School of Materials Sciences \& Engineering,
Beijing Institute of Technology, Beijing 100081, China

\textsuperscript{2}School of Computer Science and Technology, Beijing
Institute of Technology, Beijing 100081, China

e-mail: hzzhong@bit.edu.cn; fuying@bit.edu.cn

\textbf{Abstract}

Nanocrystal synthesis has been highly dependent on trial-and-error due
to the complex correlation between synthesis parameters and
physicochemical properties. Although deep learning offers a potential
methodology to achieve generative inverse design, it is still hindered
by the scarcity of high-quality datasets that align nanocrystal
synthesis routes with their properties. In this work, we develop
NanoExtractor, a large language model (LLM) with well-designed data
augmentation strategies to extract structured synthesis routes and
corresponding properties from unstructured literature. NanoExtractor
achieves a weighted average score of 92\% on the test set, significantly
outperforming other chemistry-specialized (9\%) and general-purpose LLMs
(57\%). With this model, we construct a large-scale Nanocrystal
Synthesis-Property (NSP) database containing nearly 160,000 aligned
entries. Based on this database, we further develop NanoDesigner, an LLM
for generative inverse synthesis design, achieving an F1 score of 0.85.
The applicability of NanoDesigner is experimentally validated across
multiple nanocrystal systems including MgF\textsubscript{2},
CsPbBr\textsubscript{3}, and PbS. Notably, NanoDesigner recommends a
critical non-stoichiometric precursor concentration for synthesizing
MgF\textsubscript{2} nanocrystals, which experimentally proved essential
for suppressing byproduct formation. In all, our work bridges the gap
between unstructured literature and data-driven synthesis, providing a
human-AI collaborative paradigm for accelerating material discovery.


\textbf{Introduction}

Colloidal nanocrystals are an important class of nanomaterials with
applications ranging from biomedicine to
optoelectronics\textsuperscript{1-3}, some of which have already been
applied in commercial products\textsuperscript{4,5}. The
industrialization of nanocrystals requires materials that simultaneously
satisfy multiple metrics, such as high quantum yield, precise emission
peak, and long-term stability\textsuperscript{4,6-9}. These properties
are closely related to atomic arrangements, which are fundamentally
determined by the nucleation and growth
processes\textsuperscript{10-12}. However, due to the high sensitivity
of nanocrystals to synthesis parameters and the lack of quantitative
theoretical descriptions\textsuperscript{13,14}, synthesis optimization
remains heavily reliant on labor-intensive trial-and-error exploration
of a high-dimensional parameter space\textsuperscript{15-17}.

Data-driven inverse synthesis design offers a promising solution to this
issue\textsuperscript{18,19}. In contrast to inverse design for crystal
structures from properties\textsuperscript{20,21}, inverse synthesis
design aims to generate precise synthesis routes including quantitative
reactants and conditions, customized to specific target properties.
However, given the complexity of the chemical synthesis space, achieving
effective inverse design requires massive datasets where synthesis
routes are rigorously aligned with product properties.

Existing datasets related to chemical synthesis are typically collected
through automated laboratories\textsuperscript{22,23} or text mining via
conventional natural language processing\textsuperscript{24,25}. These
datasets have been applied to predict nanocrystal sizes and optical
properties directly from synthesis recipes\textsuperscript{26-33} and to
recommend precursors\textsuperscript{34,35}. However, their utility for
generative inverse design is constrained by the scarcity of large-scale
data that aligns synthesis routes with product properties. LLMs have
revolutionized data collection with their impressive contextual
understanding and logical reasoning capabilities, presenting a unique
opportunity for constructing structured databases from
literature\textsuperscript{36-38}.

In this work, we develop NanoExtractor, an LLM dedicated to structured
information extraction. Enabled by well-designed augmentation
strategies, NanoExtractor achieves a weighted average score of 92\% on
the test set, exceeding the performance of other chemistry-specialized
(9\%) and general-purpose LLMs (57\%). This model is employed to extract
synthesis routes and corresponding product properties from the
literature, constructing an aligned NSP database. The resulting NSP
database contains approximately 160,000 aligned entries, covering
synthesis methods for a wide range of nanocrystals and nanocomposites.
Based on the NSP database, we further develop NanoDesigner for the
generative inverse design of nanocrystals. Given the target product,
specified reactants, and desired properties, NanoDesigner generates
specific candidate synthesis routes with an F1 score of 0.85.
Experimental results confirm that the model successfully generates
viable synthesis routes for CsPbBr\textsubscript{3}, PbS and PbSe
nanocrystals, as well as the rarely reported MgF\textsubscript{2}
nanocrystals.

\textbf{Results and Discussion}

\textbf{Data Annotation}

Figure 1 illustrates the construction workflow of the NSP database. Text
and tabular information are extracted from approximately 170,000
articles related to nanocrystal synthesis. A pre-trained paragraph
classifier is then employed to identify target paragraphs containing
descriptions of nanocrystal synthesis and properties. These target
paragraphs are subsequently fed into NanoExtractor to achieve the
alignment of synthesis routes with product properties, resulting in the
structured NSP database. The paragraph classifier is designed to
distinguish target paragraphs (those describing synthesis methods, size,
morphology, absorption spectra, and emission spectra) from non-relevant
text (an annotation example is shown in Figure S1) and achieves a high
recall of 0.96. An analysis of token importance within the target
paragraphs reveals that numerical values and operational verbs
associated with synthesis are critical distinguishing features (Figure
S2). These target paragraphs are annotated with synthesis steps,
synthetic routes, and product properties to construct the NanoExtractor
dataset, as shown in Figure S3a. Specifically, synthesis steps are
defined as concise sentences containing a single operational verb;
synthesis routes are sequences composed of different synthesis steps;
and product properties including size, morphology, and emission peak
positions (an annotation example is shown in Figure S3b). It is critical
that synthesis routes and product properties are linked through specific
product names.

\includegraphics[width=1.0\textwidth]{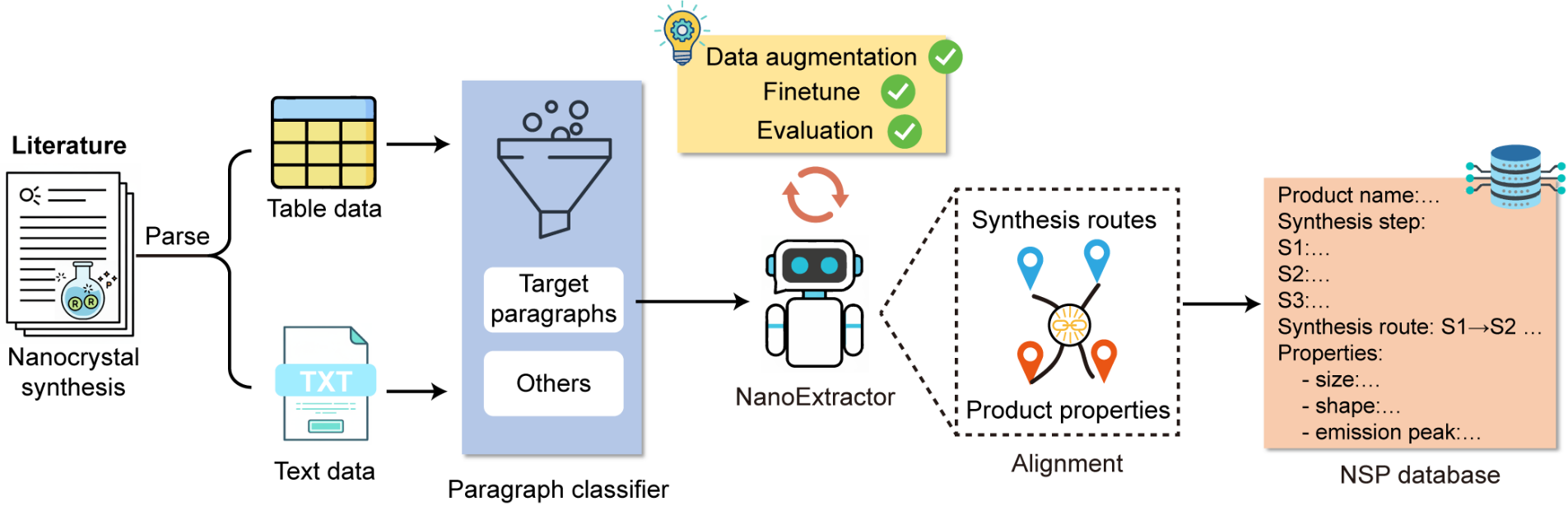}

\textbf{Figure 1 \textbar{}} \textbf{The construction workflow of the
NSP database.} Literature is first parsed into text and tables, followed
by a pre-trained classifier to identify target paragraphs. NanoExtractor
then extracts aligned synthesis routes and product properties from these
paragraphs to construct the structured NSP database.

\textbf{Data Augmentation for NanoExtractor}

To improve the performance of NanoExtractor, we propose four data
augmentation strategies targeting common failure modes in LLM-based
synthesis route extraction. First, as shown in Figure 2a,
general-purpose LLMs (e.g., DeepSeek and GPT) are utilized to rewrite
target paragraphs and corresponding synthesis steps via prompt
engineering, followed by manual verification to generate rephrased
labels. Second, to learn the error-correction capability of the model,
incorrect extraction answers are constructed by controlled exchanging,
deleting, or fabricating steps, numbers, routes, and properties (see
Figure S3c for details), which serve as negative samples during
training. Third, to mitigate model hallucinations, we generate negative
answers by replacing target paragraphs with other paragraphs and
populating the extraction fields with "NOT MENTION", thereby suppressing
extraction from non-target text. Fourth, a confidence calibration
strategy is implemented by appending low-confidence tags to labels
containing incorrect answers, while attaching high-confidence tags to
the remaining labels. This enables NanoExtractor to simultaneously
output confidence scores for its responses.

To effectively integrate both the raw data and the above augmented
samples into a unified training framework, we design two prompt
templates to simultaneously utilize both raw and augmented data (Figure
2b). Prompt \#1 instructs the model to strictly extract synthesis routes
and properties verbatim from target paragraphs, prohibiting any
inference or fabrication. Prompt \#2 permits the model to reference
incorrect answers to learn error correction. The target paragraph
followed by the prompt serves as the input, with the correct answer
(high-confidence tags) and incorrect answer (low-confidence tags) as the
output. Notably, in Prompt \#2, the incorrect answer is also included in
the input (following the target paragraph), training the model to
correct mistakes.

\includegraphics[width=1.0\textwidth]{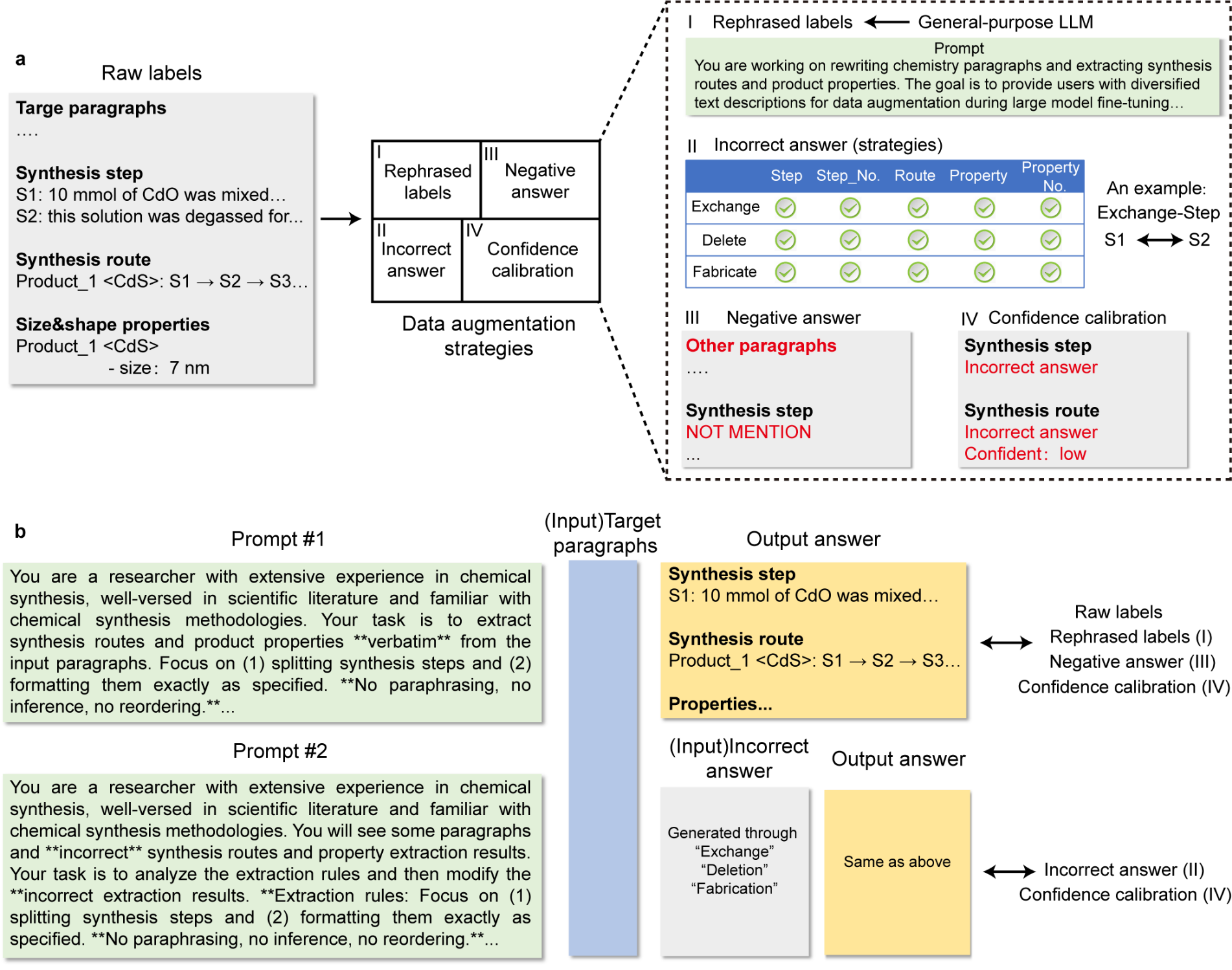}

\textbf{Figure 2 \textbar{} Data augmentation strategies and prompt
design of NanoExtractor.} (a) Schematic diagram of four data
augmentation strategies for raw labels. (b) Two prompt templates
designed for training with raw labels and four types of augmented data.

\textbf{Evaluation of NanoExtractor}

We develop a test set consisting of diverse samples, covering challenges
such as implicit operating conditions, continuous product processing and
characterization workflows, branching variables in multi-parameter
reactions, and long-context dependencies. Figures 3a and 3b show the
output of NanoExtractor and the corresponding reference output for a
representative sample from the test set (Test set\_1), respectively. The
model can well reproduce the ground truth. For instance, it correctly
identifies the reaction type in synthesis step (S1), where the product
name is a valid synonymous substitution. We invite human experts to
evaluate the model's performance according to the established scoring
criterion (see Supplementary Note 1 for details). The scoring criterion
is reference-based, with the total score computed by comparing model
predictions against the ground-truth answers. Specifically, a correct
route synthesis is awarded 10 points, with an additional 2 points earned
for each correctly identified property. A synthesis route is considered
correct only if all numerical values and operational verbs within each
step are accurate, with no omissions or redundancies. Based on this
evaluation metric, NanoExtractor achieves a score of 100\% for the
sample in Figure 3a, while another test sample yields a score of 84.2\%
(Figure S5). Figure 3c shows the weighted average scores on the test set
across different training epochs. The model achieves a peak weighted
average score of 92\% after two epochs, and extended training leads to
overfitting. Notably, training without data augmentation yields a
weighted average score of only 20\%. To ensure the reliability of these
human-evaluated scores, three independent domain experts assess the test
set, yielding an average pairwise Cohen's $\kappa$ of 0.879, which indicates
almost perfect inter-rater agreement.

Table S1 details the specific reasons for score deductions of
NanoExtractor. The model trained with two epochs loses points only due
to missing product properties and routes, while the model trained for
one epoch loses points due to misalignment between the product
properties and the synthesis route. Omission errors are acceptable in
database construction, whereas misalignment errors represent mismatches
between synthesis routes and product properties, resulting in reduced
database credibility. A t-test is used to evaluate the association
between the model's output confidence and its performance scores. As
shown in Figure S6, there is a statistically significant difference (p
\textless{} 0.05) between the scores of the high-confidence and
low-confidence groups.

To further benchmark NanoExtractor against the state-of-the-art LLMs, we
evaluate five representative models on the same test set, including both
chemistry-specialized LLMs (ChemDFM\textsuperscript{39},
ChemLLM\textsuperscript{40}, SciLitLLM\textsuperscript{41}) and advanced
general-purpose LLMs (GPT-5.2, Grok-4). As shown in Figure 3d,
NanoExtractor significantly outperforms all compared models. The
chemistry-specialized models struggle to handle the complex extraction
tasks. The weighted average scores for ChemDFM, ChemLLM, and SciLitLLM
are 3\%, 1\%, and 9\%, respectively. General-purpose models perform
better but remain insufficient for precise database construction, with
GPT-5.2 and Grok-4 scoring 57\% and 56\%, respectively. An example of an
output from chemistry-specialized LLMs and general-purpose LLMs on the
test set is provided in Supplementary Note 2.

\includegraphics[width=1.0\textwidth]{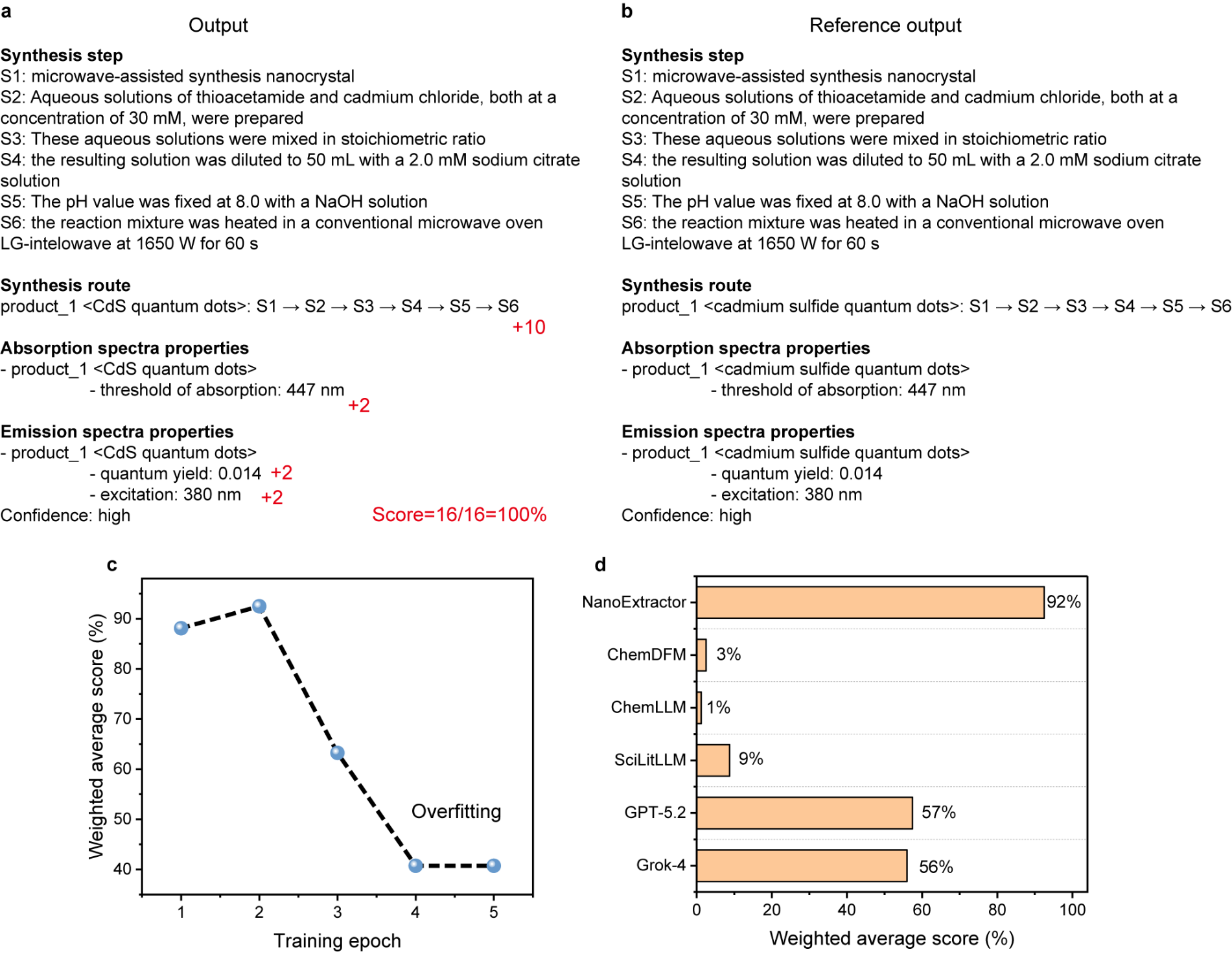}

\textbf{Figure 3 \textbar{} Performance evaluation of NanoExtractor.}
(a) NanoExtractor output for the test set sample (Test set\_1) and (b)
the reference output. (c) Weighted average test set scores evaluated by
human experts across varying training epochs. (d) Comparison of weighted
average scores on the test set for NanoExtractor against
chemistry-specialized (ChemDFM, ChemLLM, SciLitLLM) and general-purpose
(GPT-5.2, Grok-4) LLMs.

\textbf{Statistical overview of the NSP database}

Using NanoExtractor, approximately 130,000 literature sources are
extracted, and samples containing "NOT MENTION" or those with low
confidence are filtered out. As shown in Figure S7, the NSP database
contains approximately 160,000 structured synthesis routes
(corresponding to about 47,000 articles). We further evaluate a subset
of samples from the database that are excluded from both the training
and test sets, which receive high scores of 100\% (see Supplementary
Note 3). As shown in Figure 4a, the NSP database covers a wide variety
of synthetic methods for nanocrystals, including hydrothermal synthesis,
hot-injection synthesis, and heat-up synthesis, among others.
Furthermore, the database records various product properties, with a
primary focus on size and optical properties (Figure 4b). Figure 4c
shows partial statistics on the product names and the number of
corresponding synthesis routes (excluding composite and core-shell
structures). Taking CsPbBr\textsubscript{3} nanocrystals as an example,
we analyze the probability of specific reactant combinations. As
indicated in Figure 4d, the combination of PbBr\textsubscript{2},
oleylamine, and Cs\textsubscript{2}CO\textsubscript{3} occurs with a
frequency of 95\%, while toluene, hexane, ethyl acetate, and
N,N-dimethylformamide serve as the most common solvents and
antisolvents.

\includegraphics[width=1.0\textwidth]{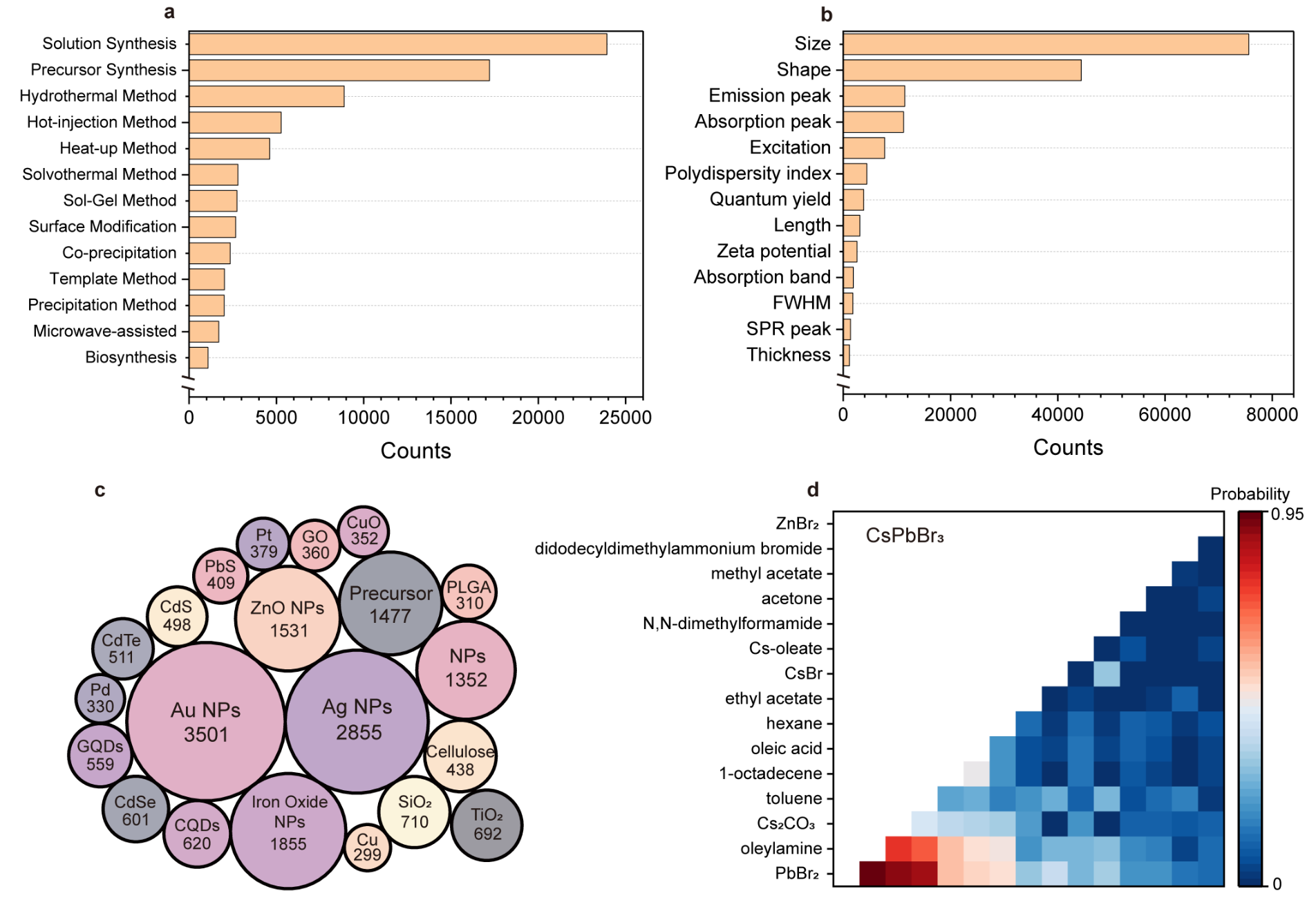}

\textbf{Figure 4} \textbf{\textbar{} Statistical overview of the NSP
database.} Partial statistics on (a) reaction types, (b) product
properties, and (c) product names recorded in the NSP database. (d)
Probability of reactant combinations for CsPbBr\textsubscript{3}
nanocrystals in the NSP database.

\textbf{Inverse design with NanoDesigner}

To demonstrate the potential of the NSP database for inverse synthesis
design of nanocrystals, we develop NanoDesigner. As shown in Figure 5a,
by inputting the target product, constrained reactants, and desired
properties (see Supplementary Note 4 for prompts), NanoDesigner is
capable of generating multiple candidate synthesis routes. We introduce
a series of evaluation metrics, including F1 scores, ROUGE
(Recall-Oriented Understudy for Gisting Evaluation), and success rates
of synthesis, to evaluate NanoDesigner from both computational and
experimental perspectives (see the Methods section for definitions). The
F1 score evaluates the model's precision in incorporating constrained
reactants, while the ROUGE score quantifies the overall sequence overlap
with the reference outputs. As shown in Figure S8, NanoDesigner exhibits
optimal performance after three fine-tuning epochs at a temperature of
0.8, with an F1 score of 0.85 and a ROUGE score of 0.42. In contrast,
the baseline achieves an F1 score of 0.95 but an extremely low ROUGE
score of 0.07, indicating a poor sequence overlap with the ground truth.

Taking the rarely reported synthesis of MgF\textsubscript{2}
nanocrystals as an example, we constrain the reactants to
MgCl\textsubscript{2} and NaF with a target size of 10 nm. NanoDesigner
proposes three distinct synthesis routes (Figure 5b and Supplementary
Note 5). Notably, literature typically reports the use of hydrofluoric
acid for synthesizing MgF\textsubscript{2}
nanocrystals\textsuperscript{42}. We intend to explore potential routes
for synthesizing MgF\textsubscript{2} nanocrystal using NaF (a safer
reactant). This requires increased generalizability of NanoDesigner
because this synthesis route does not exist in the training set.
Surprisingly, the synthesis route proposed by NanoDesigner provides
precise synthesis details, including reactant molarities, solvent
volumes, reaction temperatures, and post-processing protocols (Figure
5b). However, noting that the maximum solubility of NaF in water (0.1 M)
is lower than the concentration recommended by the model (1 M), we
adjust the precursors to c(MgCl\textsubscript{2}) = c(NaF) = 0.1 M for
the experiment while maintaining other conditions. Figure 5c shows the
experimental results, including photographs of three forms of
MgF\textsubscript{2} nanocrystals: colloidal, ethanol dispersion, and
dried colloidal. The transmission electron microscope (TEM) image of the
resulting colloidal MgF\textsubscript{2} nanocrystals is shown in Figure
5d, with an average diameter of 16.3 nm. Surprisingly, both routes
recommended by NanoDesigner suggest a non-stoichiometric molar ratio of
MgCl\textsubscript{2} to NaF (1:1), deviating from conventional chemical
intuition. We investigate the product composition at a stoichiometric
1:2 molar ratio. X-ray diffraction (XRD) analysis indicates that this
yields a mixture of MgF\textsubscript{2} and NaMgF\textsubscript{3}
(Figure S9).

We test the model's capability to propose new synthesis routes for
CsPbBr\textsubscript{3} and PbS nanocrystals to further demonstrate its
generalizability (see the description of the training set in
Supplementary Note 6). When constraining the use of bromoacetophenone as
a precursor to synthesize CsPbBr\textsubscript{3} nanocrystals, the
synthesis routes recommended by NanoDesigner include a successful route
achieving an emission peak of 517 nm for CsPbBr\textsubscript{3} (see
the first route in Supplementary Note 7 and Figure S10) and two failed
routes (see the second and third routes in Supplementary Note 7). When
OLA-S was used as the precursor to synthesize spherical PbS nanocrystals
with target sizes of 6 nm, 8 nm, and 10 nm (Supplementary Note 8), the
synthesis routes recommended by NanoDesigner are all successful with
relative size errors of 3.3\%, 7.5\%, and 14\%, respectively, as shown
in Figure S11.

We also validate the inverse design capabilities using well-established
PbSe nanocrystals, specifying PbO and tri-n-octylphosphine as reactants,
with a target size of 10 nm and spherical morphology (Supplementary Note
9). The XRD pattern and TEM image (Figure 5e and 5f) confirm the
successful synthesis of PbSe nanocrystals. Size distribution statistics
confirm an average diameter of 10.5 nm, with a relative error of 5\%
compared to the target value of 10 nm. All experimental metrics and the
reasons for failure routes are summarized in Table S2.

Moreover, the same inverse design example of MgF\textsubscript{2} is
assigned to chemistry-specialized LLMs and general-purpose LLMs. As
shown in Supplementary Note 10, ChemDFM and ChemLLM produce disorganized
responses. SciLitLLM provides an over-simplified synthesis route and
recommends the standard stoichiometric 1:2 ratio (Mg:F), which
inevitably results in impure phases. Similarly, both GPT-5.2 and Grok-4
recommend the standard stoichiometric 1:2 ratio consistent with chemical
intuition, but Grok-4's proposed room-temperature reaction is
insufficient for crystallization.

\includegraphics[width=1.0\textwidth]{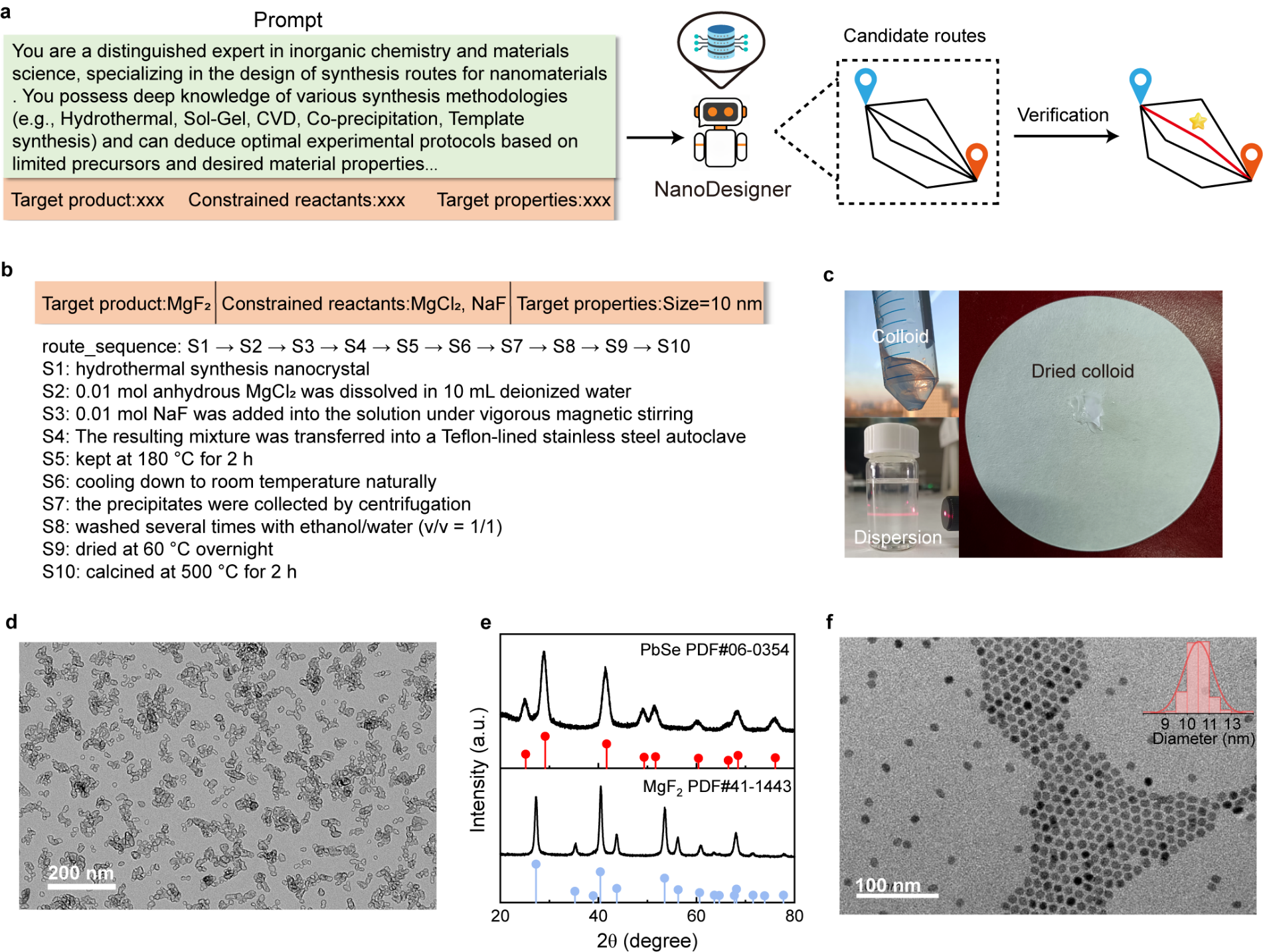}

\textbf{Figure 5} \textbf{\textbar{} Generative inverse design and
experimental validation.} (a) Schematic diagram of inverse design for
nanocrystals using NanoDesigner with the NSP database as training data.
(b) The suggested synthesis route for MgF\textsubscript{2} nanocrystals
by NanoDesigner. (c) Photographs of MgF\textsubscript{2} nanocrystals as
colloids, ethanol dispersions, and dried colloids. (d) TEM image of
MgF\textsubscript{2} nanocrystals. (e) XRD patterns of PbSe and
MgF\textsubscript{2} nanocrystals. (f) TEM images of PbSe nanocrystals,
the inset shows size distribution statistics.

\textbf{Conclusions and Outlook}

In summary, we construct a large-scale, aligned Nanocrystal
Synthesis-Property (NSP) database containing nearly 160,000 entries to
enable generative inverse design. This database is constructed using
NanoExtractor, a large language model (LLM) fine-tuned for structured
information extraction from scientific literature. By implementing four
data augmentation strategies, NanoExtractor significantly improves
extraction accuracy, mitigates hallucinations, and enables
self-assessment capabilities. Notably, compared to the model without
data augmentation, the weighted average score improves from 20\% to
92\%. This performance significantly outperforms the state-of-the-art
LLMs. The benchmark results reveal that chemistry-specialized and
general-purpose LLMs achieve weighted average scores of only 9\% and
57\%, respectively, on the same test set, underscoring the necessity of
our domain-specific fine-tuning for complex information extraction.
Using approximately 160,000 synthesis routes from the NSP database, we
further develop NanoDesigner, an LLM for the inverse synthesis design of
nanocrystals. Given specified constraints (such as the target product
and reactants), the model generates detailed synthesis routes that are
subsequently validated experimentally. Most remarkably, for the
synthesis of MgF\textsubscript{2}, the model recommends a
counter-intuitive non-stoichiometric precursor ratio, which is
experimentally confirmed to be critical for suppressing the formation of
the NaMgF\textsubscript{3} byproduct. In comparison, the
state-of-the-art LLMs (GPT and Grok-4) rely on conventional chemical
intuition and fail to identify this critical synthesis condition,
further validating the advantage of learning from a large-scale aligned
database.

Although NanoDesigner demonstrates a strong capability in generating new
synthesis routes for nanocrystals, the limitations of the model can be
summarized in three aspects. i) The reliability of generated routes
correlates with the availability of training set. For instance, the
routes of well-established PbS and PbSe with common precursors are
highly reliable (100\% success rate). In comparison, zero-shot systems
like MgF\textsubscript{2} and CsPbBr\textsubscript{3} have a lower
success rate of 66\%. ii) The model lacks a precise understanding of
solubility limits, occasionally resulting in unfeasible routes. iii) The
model has limited capability to generate complex, multi-step synthesis
routes for advanced nanocrystal structure, such as core-shell quantum
dots.

We believe the NSP database serves as a foundation for developing
forward prediction and inverse design models. Given the rapid
advancements in inverse synthesis design, it is worth developing more
refined design algorithms and integrating named entity recognition
technologies to realize the efficient and precise discovery and
optimization of nanocrystals in the future.

\textbf{Methods}

\textbf{Chemicals}

All commercially available chemicals were used without further
purification. Anhydrous magnesium chloride (MgCl\textsubscript{2}, 99\%,
3AMaterials), sodium fluoride (NaF, 99\%, 3AMaterials), deionized water
(laboratory-made), lead(II) oxide (PbO, 99.9\%, Aladdin), selenium
powder (Se, 99.99\%, Aladdin), oleic acid (OA, 90\%, Aladdin),
trioctylphosphine (TOP, 90\%, Aladdin), 1-octadecene (ODE, 90\%,
Aladdin), oleylamine (OLA, 80-90\%, Aladdin), sulfur (S, $\geq$99.95\% metals
basis, Aladdin), cesium carbonate
(Cs\textsubscript{2}CO\textsubscript{3}, $\geq$99\%, Aladdin),
2-bromoacetophenone ($\geq$98\%, Aladdin), acetone ($\geq$99.9\%, Sigma-Aldrich),
hexane ($\geq$99\%, Sigma-Aldrich), toluene (analytical-grade, Beijing
TongGuang Fine Chemicals) and ethanol (anhydrous, $\geq$99.5\%,
Sigma-Aldrich) were used in nanocrystals synthesis, purification, and
washing.

\textbf{Synthesis and Characterization of MgF\textsubscript{2}, PbSe,
PbS and CsPbBr\textsubscript{3} Nanocrystals}

\emph{MgF\textsubscript{2} nanocrystals} The synthesis of
MgF\textsubscript{2} nanocrystals followed the route suggested by the
model (Figure 5b), except that the reaction was conducted in a 100 mL
Teflon-lined stainless steel autoclave with 60 mL of deionized water
(c(MgCl\textsubscript{2}) = c(NaF) = 0.1 M), while all other conditions
remained unchanged.

\emph{PbSe nanocrystals} The synthesis of PbSe nanocrystals followed the
route suggested by the model (see the first route in Supplementary Note
9), except that the reaction was conducted in a 100 mL three-neck flask,
and all reactant quantities were scaled up 20-fold to ensure sufficient
product yield.

\emph{CsPbBr\textsubscript{3} nanocrystals} The synthesis of
CsPbBr\textsubscript{3} nanocrystals followed the route suggested by the
model (see the first route in Supplementary Note 7), except that the
reaction was conducted in a 100 mL three-neck flask, and all reactant
quantities were scaled up 5-fold to ensure sufficient product yield.

\emph{PbS nanocrystals} The synthesis of PbS nanocrystals followed the
route suggested by the model (Supplementary Note 8), except that all
reactant quantities were scaled up 2-fold to ensure sufficient product
yield.

\emph{Characterization} All nanocrystals were not subjected to any size
selection prior to TEM characterization. The nanocrystal samples were
dispersed in hexane or ethanol and added dropwise to an ultrathin
carbon-supported film (300 mesh) at 60 °C. TEM observations were
performed using a FEI Tools F200S field-emission transmission electron
microscope (FEI Co., USA) operated at 200 kV. XRD patterns were recorded
by a Bruker D8 FOCUS advance X-ray diffractometer operated at 40 kV and
200 mA current under Cu K$\alpha$ radiation (wavelength of 1.5418 Å).
Ultraviolet-visible (UV-vis) absorption spectra were measured using a
UV-6100 UV-vis spectrophotometer (Shanghai Mapada Instruments Co.,
Ltd.), and photoluminescence (PL) spectra were obtained using an F-380
fluorescence spectrometer (Tianjin Gang dong Science and Technology
Development Co., Ltd.).

\textbf{Literature Collection and Preprocessing}

Relevant article DOIs were identified by querying the CrossRef database
using keywords such as "nanomaterials", "nanocrystals", "nanoparticles",
and "quantum dots". The full-text articles were primarily acquired via
the application programming interfaces (APIs) of major publishers,
specifically Elsevier and Springer Nature. To ensure strict compliance
with copyright regulations and data usage policies, we utilized
authorized text and data mining protocols. Content was downloaded in
structured XML or HTML formats to facilitate accurate parsing. All data
acquisition was conducted in accordance with ethical guidelines, with
explicit permissions or API keys obtained from the respective publishers
to sanction the usage of their content for research purposes. Following
a rigorous data cleaning process, which involved the exclusion of review
articles, non-research content, and incomplete texts, a final corpus of
approximately 170,000 articles was retained to serve as the input for
the paragraph classifier.

\textbf{Paragraph Classifier Development}

To efficiently filter relevant text from the massive corpus, we
developed a binary classification model based on the RoBERTa-base
architecture\textsuperscript{43}. The annotated dataset was split into
training, validation, and test sets using stratified sampling to
maintain the consistency of label distribution. To address the inherent
class imbalance between target and other paragraphs, we calculated a
positive class weight based on the training set statistics and
integrated it into the binary cross entropy with logits loss function.
Rather than using a default classification threshold of 0.5, we
implemented a dynamic threshold optimization strategy. After each epoch,
the decision threshold was tuned on the validation set to maximize the
F1 score, ensuring the optimal trade-off between precision and recall.
The best-performing model configuration and its corresponding threshold
were then applied to the test set for final evaluation. The optimal
threshold is 0.26 after testing.

\textbf{Data Augmentation Strategies}

We implemented four distinct data augmentation strategies applied to the
raw labels to enhance model robustness. First, the general-purpose LLM
was utilized to rephrase the target paragraphs and extracted content
within the raw labels via prompt engineering. Each raw label was
rephrased 2\textasciitilde3 times, followed by rigorous manual
verification to ensure the quality and semantic consistency of the
augmented data. Second, to train the model's error-correction
capabilities, we constructed negative samples containing specific types
of errors. We defined 15 permutation types derived from 3 operations
(exchange, delete, and fabricate) applied to 5 target entities
(synthesis steps, numerical values within steps, route sequences,
property names, and numerical values of properties), as illustrated in
Figure 2a. The definitions of these target entities are detailed in
Figure S3b. Figure S3c shows three examples of these permutations. For
instance, the "exchange-step" operation involves exchanging the content
of synthesis steps while maintaining their original sequence numbering.
These negative samples were generated programmatically to ensure
randomness. Third, to suppress model hallucinations and prevent forced
extraction from irrelevant text, we constructed "negative answer"
labels. In these samples, the target paragraph in a raw label was
replaced with a non-relevant paragraph, and all extraction fields were
populated with "NOT MENTION". Fourth, a confidence tag was appended to
all raw and augmented labels. Specifically, labels containing incorrect
answers (from the negative sampling strategy) were tagged with a
"Confidence: low" marker at the end of the sequence, whereas all other
labels were appended with a "Confidence: high" marker. This strategy
enables the model to output a confidence assessment simultaneously with
its extraction. Figure S4a shows the quantitative distribution of the
raw and augmented data. Figure S4b shows the token count distribution of
the training samples (including prompt tokens).

\textbf{Model Training}

\emph{NanoExtractor} For the task of extracting structured
synthesis-property relationships from literature, we developed
NanoExtractor by fine-tuning the Qwen3-14B model using the LLaMA-Factory
framework. A total of 161 manually annotated samples were used for
training, and 33 held-out samples (approximately 20\% of the training
set) were used as the test set. To balance computational efficiency with
model performance, we utilized Low-Rank Adaptation (LoRA) technology.
The LoRA rank was set to 12, and the scaling factor was set to 24,
targeting all linear layers within the transformer blocks. A dropout
rate of 0.05 was applied to the LoRA layers to prevent overfitting. The
training process was optimized using the AdamW optimizer with a cosine
learning rate scheduler. The initial learning rate was set to
4×10\textsuperscript{-5} with a warmup ratio of 0.1. To accommodate the
long-context requirements of scientific literature, the maximum sequence
length (cutoff length) was set to 8,192 tokens. The model was trained in
BFloat16 precision. For the training setup, we used a per-device batch
size of 8 with gradient accumulation steps set to 2. The model was
trained for up to 5 epochs. 10\% of the training set was reserved as a
validation set. During the inference phase for information extraction,
the temperature and top-p parameters were set to 0.2 and 0.8,
respectively.

\emph{NanoDesigner} To enable the generative inverse design of
nanocrystals, we developed NanoDesigner by full fine-tuning on the
lightweight Qwen3-0.6B model. The training and validation sets do not
include the route for synthesizing MgF\textsubscript{2} nanocrystals
using NaF. The routes for synthesizing PbS nanocrystals using OLA-S and
synthesizing CsPbBr\textsubscript{3} nanocrystals using
bromoacetophenone were manually removed, while the route for PbSe
nanocrystals remains unchanged (see Supplementary Note 6 for details).
Keep 10\% of the dataset as the validation set, and leave the rest as
the training set. The training configuration included a learning rate of
3×10\textsuperscript{-5}, a per-device batch size of 8, and a gradient
accumulation step of 8 to stabilize the training updates. The maximum
sequence length was set to 2,048 tokens, which was sufficient to cover
the context of synthesis route generation. Similar to the extraction
model, we used a cosine learning rate scheduler with a 0.1 warmup ratio
and trained for 5 epochs. For the inverse design inference, to encourage
diversity and creativity in the generated synthesis routes, the sampling
parameters were adjusted to a higher temperature of
0.8\textasciitilde0.95 and a top-p value of 0.7. All experiments were
conducted on a server equipped with an NVIDIA RTX PRO 6000 GPU.

\textbf{Evaluation Metrics}

\emph{NanoExtractor} We evaluated structured extraction performance
using a reference-based scoring metric designed to reflect the
correctness of complete synthesis routes and their associated product
properties. Rather than evaluating individual tokens, our metric
operates at the level of synthesis steps, routes, and properties, which
aligns with the practical requirements of database construction. The
scoring metric was established based on a weighted system analogous to
recall, where the total score for a sample is calculated against the
reference ground truth. Detailed scoring criteria are provided in
Supplementary Note 1. The evaluation follows a hierarchical procedure,
first assessing the synthesis route, followed by the product properties.
A synthesis route is considered correct (+10 points) only if the
synthesis steps and the sequence of route perfectly match the reference.
Within each step, all numerical values must be exact matches, while
operational verbs and reaction types are evaluated based on semantic
equivalence (accepting synonyms). Any addition, omission, or fabrication
of steps results in a score of zero for the route. Under the premise of
a correct synthesis route, the corresponding product properties are then
evaluated. A property is deemed correct (+2 points per property) only if
the property name, numerical value, and unit exactly match the
reference. We defined a specific edge case for "partial correctness" (+5
points). This applies when the reference answer outlines an independent
synthesis route for a precursor, whereas the model's output correctly
merges the precursor synthesis and the final product synthesis into a
single continuous route without any omission, addition, or fabrication
of information. In such cases, the route is awarded 5 points, and the
subsequent property evaluation proceeds as normal.

To ensure the reliability of the manual scoring process, three
independent domain experts were invited to conduct a blind evaluation of
the test set based on the criteria. The pairwise unweighted Cohen's $\kappa$
for all three possible expert pairs and their average were calculated.
The $\kappa$ coefficient is calculated using the following equation:

\[\kappa = \frac{p_{o} - p_{e}}{1 - p_{e}}\]

where p\textsubscript{o} is the proportion of samples where two raters
assigned the exact same score, and p\textsubscript{e}\hspace{0pt} is the
expected agreement calculated based on the marginal frequencies of each
score. By penalizing random chance agreements, this metric provides a
rigorous assessment of human evaluation reliability.

\emph{NanoDesigner} To assess the performance of the generative inverse
design pipeline, we employed an evaluation framework comprising both
computational benchmarks and experimental validations. The computational
evaluation was performed using ROUGE-L (Recall-Oriented Understudy for
Gisting Evaluation) and F1 scores to assess the overall generation
accuracy on the held-out test set (10\% of the dataset). ROUGE-L
measures the structural and sequential similarity between the generated
route and the literature ground truth based on the longest common
subsequence. The F1 score specifically quantifies the model's capability
to adhere to user constraints, measuring the precision and recall of the
constrained reactants incorporated into the generated routes.

The experimental validation quantitatively assessed the practical
viability and accuracy of the routes suggested by the model based on
three criteria. i) Success Rate of Synthesis. The proportion of
successful routes out of the total model-recommended routes is defined
as the success rate of synthesis. A route is considered successful only
when the experimental results yield a pure target product (as confirmed
by XRD). ii) Relative Error. For specific numerical target properties
(such as size and emission peak wavelength), we define the relative
error between the measured property and the target property, calculated
as \(\frac{\left| V_{m} - V_{t} \right|}{V_{t}} \times 100\%\), where
V\textsubscript{m} and V\textsubscript{t} are the experimentally
measured and target values, respectively. iii) Morphology Consistency.
For non-numerical structural properties, we qualitatively assess the
consistency between the target and experimentally measured morphology of
products.

\textbf{Benchmarking}

As part of our benchmarking, we compared NanoExtractor and NanoDesigner
against five baseline large language models, including
chemistry-specialized models (ChemDFM, ChemLLM and SciLitLLM) and
general-purpose models (GPT-5.2 and Grok-4). All models were evaluated
on the same held-out test set, which was excluded from training and data
augmentation. Prompts were adapted to ensure a consistent extraction
format while strictly prohibiting inference or fabrication beyond the
provided text. For GPT-5.2 and Grok-4, web browsing and external tool
access were explicitly disabled during evaluation. The final score for
each model was computed as a weighted average across all test samples.

\textbf{Data availability}

All data are provided in the main text or Supporting Information. The
training code for all models is available at
https://github.com/ime1452/Synthesis-Properties-Database-for-Nanomaterials.
The model weights for NanoExtractor are available at
https://huggingface.co/Kai-gu/Qwen3-14B-finetune. The NSP database is
available at
https://huggingface.co/datasets/Kai-gu/Synthesis-Properties-Database-for-Nanomaterials.

\textbf{Supporting Information}

Additional figures, tables and model details about prompts, benchmarks,
and outputs that provide detailed insights into various aspects of the
study, including summary of reasons for losing scores (Table S1);
summary of experimental testing for NanoDesigner (Table S2); an
annotated example for the paragraph classifier (Figure S1); performance
metrics of the paragraph classifier (Figure S2); the flowchart for
annotating articles (Figure S3); data distribution of the training set
(Figure S4); NanoExtractor output for the test set sample (Figure S5);
summary of scores and confidence levels (Figure S6); data volume
variations during the construction process of the NSP database (Figure
S7); ROUGE scores and F1 scores for NanoDesigner (Figure S8); XRD
pattern of MgF\textsubscript{2} nanocrystals when the molar ratio of NaF
to MgCl\textsubscript{2} is 2:1 (Figure S9); emission and absorption
spectrum of CsPbBr\textsubscript{3} nanocrystals (Figure S10); TEM
images and size distributions of PbS nanocrystals (Figure S11).

\textbf{References}

\begin{enumerate}
\def\labelenumi{\arabic{enumi}.}
\item
  Efros, A. L.; Brus, L. E. Nanocrystal Quantum Dots: From Discovery to
  Modern Development. \emph{ACS Nano} \textbf{2021}, \emph{15},
  6192-6210.
\item
  Wu, X.-g.; Jing, Y.; Zhong, H. In Situ Fabricated Perovskite Quantum
  Dots: From Materials to Applications. \emph{Adv. Mater.}
  \textbf{2025}, \emph{37}, 2412276.
\item
  García de Arquer, F. P.; Talapin, D. V.; Klimov, V. I.; Arakawa, Y.;
  Bayer, M.; Sargent, E. H. Semiconductor quantum dots: Technological
  progress and future challenges. \emph{Science} \textbf{2021},
  \emph{373}, eaaz8541.
\item
  Won, Y.-H.; Cho, O.; Kim, T.; Chung, D.-Y.; Kim, T.; Chung, H.; Jang,
  H.; Lee, J.; Kim, D.; Jang, E. Highly efficient and stable
  InP/ZnSe/ZnS quantum dot light-emitting diodes. \emph{Nature}
  \textbf{2019}, \emph{575}, 634-638.
\item
  Lin, R.; Gao, H.; Lou, J.; Xu, J.; Yin, M.; Wu, P.; Liu, C.; Guo, Y.;
  Wang, E.; Yang, S.; Liu, R.; Zhou, D.; Ding, C.; Bui, A. D.; Yin, N.;
  Macdonald, D. H.; Ma, C. Q.; Chen, Q.; Xiao, K.; Luo, X.; et al.
  All-perovskite tandem solar cells with dipolar passivation.
  \emph{Nature} \textbf{2025}, \emph{648}, 600-606.
\item
  Aqoma, H.; Lee, S.-H.; Imran, I. F.; Hwang, J.-H.; Lee, S.-H.; Jang,
  S.-Y. Alkyl ammonium iodide-based ligand exchange strategy for
  high-efficiency organic-cation perovskite quantum dot solar cells.
  \emph{Nat. Energy} \textbf{2024}, \emph{9}, 324-332.
\item
  Moon, H.; Lee, C.; Lee, W.; Kim, J.; Chae, H. Stability of Quantum
  Dots, Quantum Dot Films, and Quantum Dot Light-Emitting Diodes for
  Display Applications. \emph{Adv. Mater.} \textbf{2019}, \emph{31},
  1804294.
\item
  Wu, X.-g.; Ji, H.; Yan, X.; Zhong, H. Industry outlook of perovskite
  quantum dots for display applications. \emph{Nat. Nanotechnol.}
  \textbf{2022}, \emph{17}, 813-816.
\item
  Lee, H.; Song, H.-J.; Shim, M.; Lee, C. Towards the commercialization
  of colloidal quantum dot solar cells: perspectives on device
  structures and manufacturing. \emph{Energy Environ. Sci.}
  \textbf{2020}, \emph{13}, 404-431.
\item
  Liu, L.; Long, Z.; Shi, K.; Zhong, H. A General Crystallization
  Picture of Quantum Dots: The Underlying Physical Chemistry. \emph{CCS
  Chem.} \textbf{2025}, \emph{7}, 926-949.
\item
  Long, Z.; Liu, M.; Wu, X.-g.; Gu, K.; Yang, G.; Chen, Z.; Liu, Y.;
  Liu, R.; Zhong, H. A reactivity-controlled epitaxial growth strategy
  for synthesizing large nanocrystals. \emph{Nat. Synth.} \textbf{2023},
  \emph{2}, 296-304.
\item
  Li, S.; Du, X.; Liu, Z.; Li, Y.; Shao, Y.; Jin, R. Size Effects of
  Atomically Precise Gold Nanoclusters in Catalysis. \emph{Precis.
  Chem.} \textbf{2023}, \emph{1}, 14-28.
\item
  Horani, F.; Sharma, K.; Abu-Hariri, A.; Lifshitz, E. Colloidal Control
  of Branching in Metal Chalcogenide Semiconductor Nanostructures.
  \emph{J. Phys. Chem. Lett.} \textbf{2023}, \emph{14}, 3794-3804.
\item
  Whitehead, C. B.; Özkar, S.; Finke, R. G. LaMer's 1950 Model for
  Particle Formation of Instantaneous Nucleation and
  Diffusion-Controlled Growth: A Historical Look at the Model's Origins,
  Assumptions, Equations, and Underlying Sulfur Sol Formation Kinetics
  Data. \emph{Chem. Mater.} \textbf{2019}, \emph{31}, 7116-7132.
\item
  Braham, E. J.; Cho, J.; Forlano, K. M.; Watson, D. F.; Arròyave, R.;
  Banerjee, S. Machine Learning-Directed Navigation of Synthetic Design
  Space: A Statistical Learning Approach to Controlling the Synthesis of
  Perovskite Halide Nanoplatelets in the Quantum-Confined Regime.
  \emph{Chem. Mater.} \textbf{2019}, \emph{31}, 3281-3292.
\item
  Liu, Z.-S.; Wang, Y.; Zhao, F.; Li, H.-H.; Liu, W.-Z.; Shen, W.-S.;
  Duan, H.-W.; Wang, Y.-K.; Liao, L.-S. Liquid bidentate ligand for full
  ligand coverage towards efficient near-infrared perovskite quantum dot
  LEDs. \emph{Light Sci. Appl.} \textbf{2025}, \emph{14}, 35.
\item
  Ren, Y.; Li, C.; Fang, Y.; Pang, S.; Jiang, X.; Li, M.; Du, Z. In
  Situ, Treatment with Guanidinium Chloride Ligand Enables Efficient
  Blue Quantum Dot Light-Emitting Diodes with 23.5\% External Quantum
  Efficiency. \emph{Adv. Mater.} \textbf{2025}, \emph{37}, 2413183.
\item
  Zunger, A. Inverse design in search of materials with target
  functionalities. \emph{Nat. Rev. Chem.} \textbf{2018}, \emph{2}, 0121.
\item
  Choi, S. E.; Jang, M.; Yoon, S.; Yoo, S.; Ahn, J.; Kim, M.; Kim,
  H.-G.; Jung, Y.; Park, S.; Kim, Y.-S.; Kim, T. LLM-Driven Synthesis
  Planning for Quantum Dot Materials Development. \emph{J. Chem. Inf.
  Model.} \textbf{2025}, \emph{65}, 2748-2758.
\item
  Zeni, C.; Pinsler, R.; Zügner, D.; Fowler, A.; Horton, M.; Fu, X.;
  Wang, Z.; Shysheya, A.; Crabbé, J.; Ueda, S.; Sordillo, R.; Sun, L.;
  Smith, J.; Nguyen, B.; Schulz, H.; Lewis, S.; Huang, C.-W.; Lu, Z.;
  Zhou, Y.; Yang, H.; et al. A generative model for inorganic materials
  design. \emph{Nature} \textbf{2025}, \emph{639}, 624-632.
\item
  Ren, Z.; Tian, S. I. P.; Noh, J.; Oviedo, F.; Xing, G.; Li, J.; Liang,
  Q.; Zhu, R.; Aberle, A. G.; Sun, S.; Wang, X.; Liu, Y.; Li, Q.;
  Jayavelu, S.; Hippalgaonkar, K.; Jung, Y.; Buonassisi, T. An
  invertible crystallographic representation for general inverse design
  of inorganic crystals with targeted properties. \emph{Matter}
  \textbf{2022}, \emph{5}, 314-335.
\item
  Slattery, A.; Wen, Z.; Tenblad, P.; Sanjosé-Orduna, J.; Pintossi, D.;
  den Hartog, T.; Noël, T. Automated self-optimization, intensification,
  and scale-up of photocatalysis in flow. \emph{Science} \textbf{2024},
  \emph{383}, eadj1817.
\item
  Szymanski, N. J.; Rendy, B.; Fei, Y.; Kumar, R. E.; He, T.; Milsted,
  D.; McDermott, M. J.; Gallant, M.; Cubuk, E. D.; Merchant, A.; Kim,
  H.; Jain, A.; Bartel, C. J.; Persson, K.; Zeng, Y.; Ceder, G. An
  autonomous laboratory for the accelerated synthesis of novel
  materials. \emph{Nature} \textbf{2023}, \emph{624}, 86-91.
\item
  Mavračić, J.; Court, C. J.; Isazawa, T.; Elliott, S. R.; Cole, J. M.
  ChemDataExtractor 2.0: Autopopulated Ontologies for Materials Science.
  \emph{J. Chem. Inf. Model.} \textbf{2021}, \emph{61}, 4280-4289.
\item
  Wang, Z.; Kononova, O.; Cruse, K.; He, T.; Huo, H.; Fei, Y.; Zeng, Y.;
  Sun, Y.; Cai, Z.; Sun, W.; Ceder, G. Dataset of solution-based
  inorganic materials synthesis procedures extracted from the scientific
  literature. \emph{Sci. Data} \textbf{2022}, \emph{9}, 231.
\item
  Gu, K.; Liang, Y.; Su, J.; Sun, P.; Peng, J.; Miao, N.; Sun, Z.; Fu,
  Y.; Zhong, H.; Zhang, J. Deep Learning Models for Colloidal
  Nanocrystal Synthesis. \emph{ACS Nano} \textbf{2025}, \emph{19},
  39025-39034.
\item
  Kim, M. A.; Ai, Q.; Norquist, A. J.; Schrier, J.; Chan, E. M. Active
  Learning of Ligands That Enhance Perovskite Nanocrystal Luminescence.
  \emph{ACS Nano} \textbf{2024}, \emph{18}, 14514-14522.
\item
  Wu, Y.; Wang, C.-F.; Ju, M.-G.; Jia, Q.; Zhou, Q.; Lu, S.; Gao, X.;
  Zhang, Y.; Wang, J. Universal machine learning aided synthesis
  approach of two-dimensional perovskites in a typical laboratory.
  \emph{Nat. Commun.} \textbf{2024}, \emph{15}, 138.
\item
  Hong, Q.; Wang, X.-Y.; Gao, Y.-T.; Lv, J.; Chen, B.-B.; Li, D.-W.;
  Qian, R.-C. Customized Carbon Dots with Predictable Optical Properties
  Synthesized at Room Temperature Guided by Machine Learning.
  \emph{Chem. Mater.} \textbf{2022}, \emph{34}, 998-1009.
\item
  Rao, Z.; Tung, P.-Y.; Xie, R.; Wei, Y.; Zhang, H.; Ferrari, A.;
  Klaver, T. P. C.; Körmann, F.; Sukumar, P. T.; Kwiatkowski da Silva,
  A.; Chen, Y.; Li, Z.; Ponge, D.; Neugebauer, J.; Gutfleisch, O.;
  Bauer, S.; Raabe, D. Machine learning--enabled high-entropy alloy
  discovery. \emph{Science} \textbf{2022}, \emph{378}, 78-85.
\item
  Baum, F.; Pretto, T.; Köche, A.; Santos, M. J. L. Machine Learning
  Tools to Predict Hot Injection Syntheses Outcomes for II--VI and
  IV--VI Quantum Dots. \emph{J. Phys. Chem. C} \textbf{2020},
  \emph{124}, 24298-24305.
\item
  Nguyen, H. A.; Dou, F. Y.; Park, N.; Wu, S.; Sarsito, H.; Diakubama,
  B.; Larson, H.; Nishiwaki, E.; Homer, M.; Cash, M.; Cossairt, B. M.
  Predicting Indium Phosphide Quantum Dot Properties from Synthetic
  Procedures Using Machine Learning. \emph{Chem. Mater.} \textbf{2022},
  \emph{34}, 6296-6311.
\item
  Williamson, E. M.; Tappan, B. A.; Mora-Tamez, L.; Barim, G.; Brutchey,
  R. L. Statistical Multiobjective Optimization of Thiospinel
  CoNi\textsubscript{2}S\textsubscript{4} Nanocrystal Synthesis via
  Design of Experiments. \emph{ACS Nano} \textbf{2021}, \emph{15},
  9422-9433.
\item
  Song, Z.; Lu, S.; Ju, M.; Zhou, Q.; Wang, J. Accurate prediction of
  synthesizability and precursors of 3D crystal structures via large
  language models. \emph{Nat. Commun.} \textbf{2025}, \emph{16}, 6530.
\item
  Karpovich, C.; Pan, E.; Jensen, Z.; Olivetti, E. Interpretable Machine
  Learning Enabled Inorganic Reaction Classification and Synthesis
  Condition Prediction. \emph{Chem. Mater.} \textbf{2023}, \emph{35},
  1062-1079.
\item
  Schilling-Wilhelmi, M.; Ríos-García, M.; Shabih, S.; Gil, M. V.;
  Miret, S.; Koch, C. T.; Márquez, J. A.; Jablonka, K. M. From text to
  insight: large language models for chemical data extraction.
  \emph{Chem. Soc. Rev.} \textbf{2025}, \emph{54}, 1125-1150.
\item
  Kang, Y.; Lee, W.; Bae, T.; Han, S.; Jang, H.; Kim, J. Harnessing
  Large Language Models to Collect and Analyze Metal--Organic Framework
  Property Data Set. \emph{J. Am. Chem. Soc.} \textbf{2025}, \emph{147},
  3943-3958.
\item
  Zhang, J.; Li, J.; Zhao, G.; Wang, Q.; Guo, Y.-G.; Yang, C. Mining
  Solid-State Electrolytes from Metal--Organic Framework Databases
  through Large Language Models and Representation Clustering. \emph{J.
  Am. Chem. Soc.} \textbf{2025}, \emph{147}, 40496-40506.
\item
  Zhao, Z.; Ma, D.; Chen, L.; Sun, L.; Li, Z.; Xia, Y.; Chen, B.; Xu,
  H.; Zhu, Z.; Zhu, S.; Fan, S.; Shen, G.; Yu, K.; Chen, X. Developing
  ChemDFM as a large language foundation model for chemistry. \emph{Cell
  Rep. Phys. Sci.} \textbf{2025}, \emph{6}, 102523.
\item
  Zhang, D.; Liu, W.; Tan, Q.; Chen, J.; Yan, H.; Yan, Y.; Li, J.;
  Huang, W.; Yue, X.; Ouyang, W.; Zhou, D.; Zhang, S.; Su, M.; Zhong,
  H.-S.; Li, Y. ChemLLM: A Chemical Large Language Model. 2024,
  2402.06852. arXiv. https://arxiv.org/abs/2402.06852 (accessed Apr. 25,
  2024).
\item
  Li, S.; Huang, J.; Zhuang, J.; Shi, Y.; Cai, X.; Xu, M.; Wang, X.;
  Zhang, L.; Ke, G.; Cai, H. SciLitLLM: How to Adapt LLMs for Scientific
  Literature Understanding. 2025, 2408.15545. arXiv.
  https://arxiv.org/abs/2408.15545 (accessed Apr. 18, 2025).
\item
  Karthik, D.; Pendse, S.; Sakthivel, S.; Ramasamy, E.; Joshi, S. V.
  High performance broad band antireflective coatings using a facile
  synthesis of ink-bottle mesoporous MgF\textsubscript{2} nanoparticles
  for solar applications. \emph{Sol. Energy Mater. Sol. Cells}
  \textbf{2017}, \emph{159}, 204-211.
\item
  Liu, Y.; Ott, M.; Goyal, N.; Du, J.; Joshi, M.; Chen, D.; Levy, O.;
  Lewis, M.; Zettlemoyer, L.; Stoyanov, V. RoBERTa: A Robustly Optimized
  BERT Pretraining Approach. 2019, 1907.11692. arXiv.
  https://arxiv.org/abs/1907.11692 (accessed Jul. 26, 2019).
\end{enumerate}

\textbf{Acknowledgements}

This work is granted by the National Natural Science Foundation of China
(Nos. 52525309, 62331006 and U23A20683) and Beijing Municipal Science \&
Technology Commission, Administrative Commission of Zhongguancun Science
under Park No. Z231100006023018. We thank Ms. Yuqin Cui and Ms. Xiaoyu
Zhang for their support of computational resources, and Dr. Shipei Sun
for his assistance with material synthesis.

\textbf{Author information}

Authors and Affiliations

MIIT Key Laboratory for Low-Dimensional Quantum Structure and Devices,
School of Materials Sciences \& Engineering, Beijing Institute of
Technology, Beijing 100081, China

Kai Gu, Senliang Peng, Aotian Guo, Haizheng Zhong

School of Computer Science and Technology, Beijing Institute of
Technology, Beijing 100081, China

Yingping Liang, Ying Fu

Contributions

K. G., H. Z. and Y. F. conceived the project. K. G., S. P. and A. G.
synthesized and characterized the nanocrystals. K. G. and Y. L.
performed data cleaning, annotation, model training, and model
evaluation. K. G., Y. L., Y. F. and H. Z. analyzed the models and wrote
the manuscript.

Corresponding authors

Correspondence to Haizheng Zhong or Fu Ying

\textbf{Notes}

The authors declare no competing interests.

\textbf{Associated Content}

An initial draft of this work was uploaded to the arXiv preprint server
on January 4, 2026, as the following reference: Kai Gu; Yingping Liang;
Senliang Peng, etc. A large-scale nanocrystal database with aligned
synthesis and properties enabling generative inverse design. 2026,
2601.02424. arXiv. https://arxiv.org/abs/2601.02424 (accessed Jan. 4,
2026).

\textbf{Table of Contents}

\includegraphics[width=1.0\textwidth]{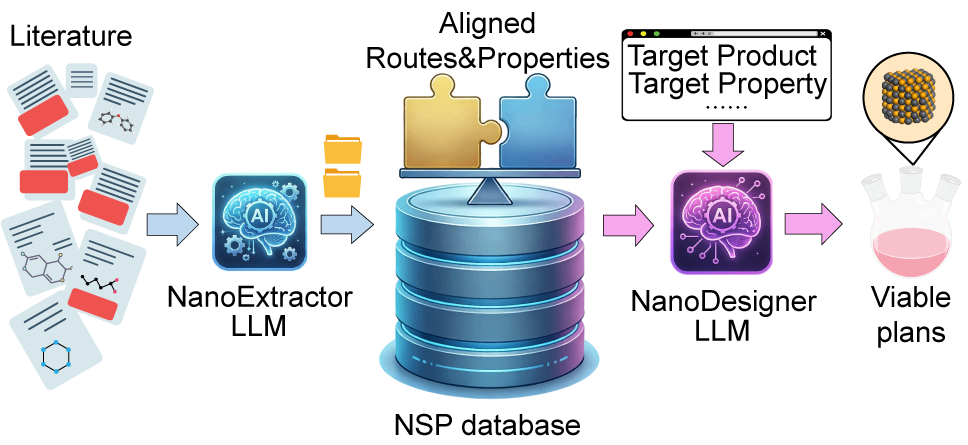}

\end{document}